\documentclass[reprint,preprintnumbers,prd,nofootinbib]{revtex4-1}

\usepackage{amsfonts}
\usepackage{amsmath}
\usepackage{array}
\usepackage{braket}
\usepackage{epstopdf}
\usepackage{graphicx}
\usepackage{hepparticles}
\usepackage{hepnicenames}
\usepackage{hepunits}
\usepackage{hyperref}
\usepackage[%
    utf8
]{inputenc}
\usepackage{slashed}
\usepackage{subfigure}
\usepackage{placeins}
\usepackage[%
    normalem
]{ulem}
\usepackage[%
    usenames,
    svgnames,
    dvipsnames
]{xcolor}

\newcommand{\ie}{\textit{i.e.}}

\newcommand{\order}[1]{\mathcal{O}\left({#1}\right)}

\newcommand{\refeq}[1]{eq.~(\ref{eq:#1})}
\newcommand{\refeqs}[2]{eqs.~(\ref{eq:#1})--(\ref{eq:#2})}
\newcommand{\reffig}[1]{figure~\ref{fig:#1}}
\newcommand{\refsec}[1]{section~\ref{sec:#1}}
\newcommand{\reftab}[1]{table~\ref{tab:#1}}

\renewcommand{\theta}{\vartheta}

\newcommand{\para}{\parallel}
\newcommand{\Gfermi}{G_F}
\newcommand{\dd}[2][]{{\mathrm{d}^{#1}}#2\,}

\let\Re\ReNew

\newcommand{\wilson}[2][]{\mathcal{C}^\text{#1}_{#2}}
\newcommand{\op}[1]{\mathcal{O}_{#1}}

\begin{document}

\allowdisplaybreaks

\preprint{EOS-2016-03, ZU-TH-12/16}
\title{Measuring the Breaking of Lepton Flavour Universality in $B\to K^*\ell^+\ell^-$}
\author{Nicola Serra}
\email{nicola.serra@cern.ch}
\author{Rafael Silva Coutinho}
\email{rafael.silva.coutinho@cern.ch}
\author{Danny van Dyk}
\email{dvandyk@physik.uzh.ch}
\affiliation{Physik-Institut, Universit\"at Z\"urich, Winterthurer Strasse 190, 8057 Z\"urich, Switzerland}

\begin{abstract}
    We propose measurements of weighted differences of the angular observables
    in the rare decays $B \to K^*\ell^+\ell^-$. The proposed observables
    are very sensitive to the difference between the Wilson coefficients $\mathcal{C}_9^{(e)}$
    and $\mathcal{C}_9^{(\mu)}$ for decays into electrons and muons, respectively.
    At the same time, the charm-induced hadronic contributions
    are kinematically suppressed to $\lesssim 7\% (4\%)$ in the region
    $1\,\GeV^2 \leq q^2 \leq 6\,\GeV^2$, as long as LFU breaking occurs only in $\wilson[$(\ell)$]{9}$. This suppression
    becomes stronger for the region of low hadronic recoil, $q^2 \geq 15\,\GeV^2$.
\end{abstract}

\maketitle

\section{Introduction}
\label{sec:intro}

In this letter, we investigate the suitability of new observables to
measure the breaking of Lepton-Flavour Universality (LFU) in rare $b\to s\ell\ell$
transitions. While measurements \cite{Aaij:2014ora,BaBarPreliminary} of the ratio $R_K$,
\begin{align}
    R_K \equiv \frac{\mathcal{B}(B \to K\mu^+\mu^-)}{\mathcal{B}(B\to K e^+e^-)}\,,
\end{align}
for dilepton masses $1\,\GeV^2 \leq q^2 \leq 6\,\GeV^2$,
hint toward LFU breaking, there has been no unambiguous discovery of such
effects. It has been proposed \cite{Hiller:2014ula} to expand such measurements
to the decays $B\to X_s\ell^+\ell^-$, $B\to K^*\ell^+\ell^-$ and $B_s\to
\phi\ell^+\ell^-$, introducing similar ratios $R_{X_s}$, $R_{K^*}$ and $R_{\phi}$,
respectively.
Analysing LFU breaking in angular observables of the decay $B \to
K^*\ell^+\ell^-$ has been proposed in \cite{Altmannshofer:2015sma}, and
more recently studied in \cite{Capdevila:2016ivx}.  Within this letter we
propose observables that can be used to accurately measure the size of this
breaking, specifically in the decays $B \to K^*\ell^+\ell^-$.  Our
study focuses on observables in which charm-induced long-distance
contributions can be kinematically suppressed.\\

The exclusive decays $B \to K^*\ell^+\ell^-$ for $\ell = e, \mu$
are governed by the effective field theory for flavour-changing neutral $b\to
s\lbrace \gamma, \ell^+\ell^-\rbrace$ transitions; see e.g. \cite{Bobeth:2012vn}.
The theory's Hamiltonian density to leading power in $G_F$ is
\begin{equation}
\begin{aligned}
  \label{eq:Heff}
  {\cal{H}}^{(\ell)}_{\rm eff}
  & = - \frac{4\, G_F}{\sqrt{2}} V_{tb}^{} V_{ts}^* \,\Bigg\lbrace\,\left[\frac{\alpha_e}{4 \pi}\,
       \sum_i \wilson[$(\ell)$]{i}(\mu)  \op{i}^{(\ell)}\right]\\
  & + \left[\frac{e}{(4\pi)^2} \sum_j \wilson{j}(\mu) \op{j}\right] + \left[\sum_k \wilson{k}(\mu) \op{k}\right]\\
  & + \order{(V_{ub} V_{us}^*) / (V_{tb} V_{ts}^*)} + \text{h.c.}\Bigg\rbrace\,,
\end{aligned}
\end{equation}
where $\wilson{i}(\mu)$ denotes the Wilson coefficients at the renormalisation
scale $\mu$, and $\op{\lbrace i,j,k\rbrace}$ denotes a basis of dimension-6
field operators.  The index $i$ iterates over all semileptonic operators,
$i=9,9',10,10',S,S',P,P',T,T5$, which are dependent on the final state lepton
flavour $\ell = e, \mu$. The indices $j$ and $k$ iterate over the radiative
($j=7,7'$), and the four-quark and chromomagnetic operators ($k=1,\dots,6,8$),
respectively. The most relevant operators read
\begin{equation}
\label{eq:SM:ops}
\begin{aligned}
    \op{7(7')} & = \frac{m_b}{e}\!\left[\bar{s} \sigma^{\mu\nu} P_{R(L)} b\right] F_{\mu\nu}\,,\\
    \op{9(9')} & = \left[\bar{s} \gamma_\mu P_{L(R)} b\right]\!\left[\bar{\ell} \gamma^\mu \ell\right]\,,\\
    \op{10(10')} & = \left[\bar{s} \gamma_\mu P_{L(R)} b\right]\!\left[\bar{\ell} \gamma^\mu \gamma_5 \ell\right]\,,
\end{aligned}
\end{equation}
where a primed index indicates a flip of the quarks' chiralities with respect
to the unprimed, SM-like operator.

Hadronic matrix elements of the semileptonic operators are parametrized in terms of
form factors, which can be determined using non-perturbative methods such as lattice
QCD (see e.g. \cite{Horgan:2013hoa}) and QCD sum rules (see e.g. \cite{Straub:2015ica}).
However, hadronic matrix elements of the correlator between four-quark operators %
$\op{i} \sim \left[\bar{s} \Gamma_i b\right]\,\left[\bar{q}\Gamma_i^\prime q\right]$, $i=1,\dots,6$
as well as the chromomagnetic operator $\op{8}$ on the one hand, and the
electromagnetic current on the other, are more complicated to estimate.  These
non-local matrix elements contribute to the transition amplitudes
$A_{\lambda}$, with $\lambda = 0,\perp,\para$, through shifts
$\wilson[$(\ell)$]{9} \mapsto \wilson[$(\ell)$]{9} + h_{9,\lambda}(q^2)$ and
$\wilson{7} \mapsto \wilson{7} + h_{7,\lambda}$.  Note that the shifts
to $\wilson{9}$ are explicitly dependent on $q^2$, the momentum transfer to the
lepton pair.

Within ratios of observables for either $\ell=\mu$ or $\ell=e$ final states,
the non-local contributions $h_{9,\lambda}(q^2)$ do not cancel completely.
However, within differences of angular observables they can be kinematically
suppressed.\footnote{%
    In reference \cite{Capdevila:2016ivx} the authors propose observables
    $Q_i$, $T_i$ and $B_i$, which are constructed from differences of the
    principal angular observables $J_i$ in $\mu$ and $e$ final states. Of the
    proposed observables, the $B_i \equiv (J_i^{(\mu)} - J_i^{(e)}) / J_i^{(e)}$
    observables are closest to what we propose here. However, their normalisation
    to the electron-mode angular observable lifts the kinematic suppression of
    the charm-induced long-distance contributions that we aim for.
}\\

The remainder of this letter is structured as follows: We propose the new
observables in \refsec{obs}. Their numerical evaluations and theoretical uncertainties are
discussed in \refsec{numerics}.  Thereafter, we study the experimental
feasibility of their measurements for both future Belle-II and LHCb data sets
in \refsec{exp}, before we conclude in \refsec{conclusion}.

\section{Measures of LFU Breaking}
\label{sec:obs}

The angular PDF for $B\to K^*(\to K \pi)\ell^+\ell^-$ decays is well known in
the literature, and we use the conventions specified in
\cite{Altmannshofer:2008dz}. There, the CP-averaged and normalized angular
observables are
\begin{equation}
    \label{eq:definition-Si}
    S_{i}(q^2) \equiv \frac{4}{3}\,\frac{J_i(q^2) + \overline{J}_i(q^2)}{\dd \Gamma/\dd q^2 + \dd {\overline{\Gamma}}/\dd q^2}\,,
\end{equation}
where unbarred quantities stem from the decay $\bar{B}\to
\bar{K}^*\ell^+\ell^-$, and the bar indicates CP conjugation.  For the
definitions of the $J_i$, see
\cite{Kruger:1999xa,Kruger:2005ep,Altmannshofer:2008dz,Bobeth:2012vn}. \footnote{%
    Note that the definition of the angular observables does not account
    for purely QED-induced modifications to the overall angular distribution;
    see \cite{Huber:2015sra,Gratrex:2015hna} for recent discussions.
}
Here
and throughout the rest of this letter, we will refer to $S_i^{(\ell)}$ and
$\Gamma^{(\ell)}$ as one of the angular observables or the decay width for the
$\ell$ final state, respectively.

All spin-averaged observables can be expressed in terms of sesquilinear
combinations of up to 14 transversity amplitudes when working in the full basis
of dimension-six semileptonic operators \cite{Bobeth:2012vn}.  For the
discussions at hand, however, we restrict our study to the operators
$\op{9,10}$.  In this case, all observables can be expressed in terms of only
$7$ transversity amplitudes,
\begin{align}
\label{eq:def-transamps}
    \frac{A_0^{L(R)}}{N \sqrt{\beta_\ell}}     & = -\left[\left(\wilson[eff,$(\ell)$]{9,0} \mp \wilson{10}\right) F_{V,0} \! + \! \frac{2 m_b}{M_B} \wilson[eff]{7,0} F_{T,0}\!\right]\\
\nonumber
    \frac{A_\para^{L(R)}}{N \sqrt{\beta_\ell}} & = -\left[\left(\wilson[eff,$(\ell)$]{9,\para} \mp \wilson{10}\right) F_{V,\para} \! + \! \frac{2 m_b M_B}{q^2} \wilson[eff]{7,\para} F_{T,\para}\!\right]\\
\nonumber
    \frac{A_\perp^{L(R)}}{N \sqrt{\beta_\ell}} & = \left[\left(\wilson[eff,$(\ell)$]{9,\perp} \mp \wilson{10}\right) F_{V,\perp} \! + \! \frac{2 m_b M_B}{q^2} \wilson[eff]{7,\perp} F_{T,\perp}\!\right]
\end{align}
as well as $A_t$. The latter is not relevant to the discussions at hand.
Note that our convention for the normalization constant $N$
\begin{equation}
    N \equiv \Gfermi \alpha_e V_{tb} V_{ts}^* \sqrt{\frac{q^2 \sqrt{\lambda}}{3\cdot 2^{10} \pi^5 M_B^3}}\,,
\end{equation}
differs from, e.g., the normalization $N^\text{\cite{Bobeth:2012vn}}$ as
used in reference \cite{Bobeth:2012vn}: $N^\text{\cite{Bobeth:2012vn}} =
\sqrt{\beta_\ell} N$. Our choice ensures that the normalization is universal
for all lepton flavours.\\

We propose to measure weighted differences of angular observables,
\begin{equation}
    \label{eq:recipe-Di}
    D_i(q^2) \equiv \frac{\dd{\mathcal{B}^{(e)}}}{\dd q^2} S^{(e)}_i(q^2) - \frac{\dd{\mathcal{B}^{(\mu)}}}{\dd q^2} S^{(\mu)}_i(q^2)\,.
\end{equation}
Assuming LFU breaking only\footnote{%
    Note that lepton-universal NP effects are not precluded here.
} %
in the Wilson coefficient $\wilson{9}$, we obtain for the indices $i=4,5,6s$ the expressions
\begin{equation}
\label{eq:results-Di}
\begin{aligned}
    \frac{-D_{4}(q^2)}{\sqrt{2}\,N^2\,\tau_B}
        & = \Re\left[\Delta_{9^2}^{(3)}\right] F_{V,\para} F_{V,0}\\
        & - \Re\left[\Delta_9^{(3)} \dots\right] + \order{\Delta_\beta^{(3)}}\,,\\
    \frac{D_{5}(q^2)}{4\,\sqrt{2}\,N^2\,\tau_B}
        & = \Re \left\lbrace \Delta_9^{(2)} \wilson{10}\right\rbrace F_{V,\perp} F_{V,0}\\
        & + \order{\Delta_\beta^{(2)}}\\
    \frac{D_{6s}(q^2)}{8\,N^2\,\tau_B}
        & = \Re \left\lbrace \Delta_9^{(2)} \wilson{10}\right\rbrace F_{V,\para} F_{V,\perp}\\
        & + \order{\Delta_\beta^{(2)}}\\
\end{aligned}
\end{equation}
where $\tau_B$ is the lifetime of the $B$ mesons, and the dots indicate
an unsuppressed expression linear in the non-local contributions $h_{9,0}(q^2)$
and $h_{9,\para}(q^2)$. Moreover, we introduce
\begin{equation}
\begin{aligned}
    \Delta_{9^l}^{(k)} & \equiv \beta_e^k \left[\wilson[$(e)$]{9}\right]^l - \beta_\mu^k \left[\wilson[$(\mu)$]{9}\right]^l\,,\\
    \Delta_\beta^{(k)} & \equiv \beta_e^k - \beta_\mu^k\,.
\end{aligned}
\end{equation}
We find that \refeq{results-Di} holds up to corrections of order $\beta_e^3 - \beta_\mu^3$ (for $D_4$)
and $\beta_e^2 - \beta_\mu^2$ (for $D_{5,6s}$). Note that $D_4$ is free of hadronic contributions
in the term $\propto \Delta_{9^2}^{(3)}$, but not free of them in the linear term $\Delta_9^{(3)}$.
For the full results, see \refeqs{results-D4-full}{results-D6s-full}.
The expressions \refeq{results-Di} hold \emph{in the entire $q^2$ spectrum},
since no explicit expression for the hadronic two-point correlation functions,
$h_{9,\lambda}(q^2) \equiv \wilson[eff,($\ell$)]{9,\lambda}(q^2) -
\wilson[$(\ell)$]{9}$, have been used. We emphasize that this also holds in
between the two vetoes for the $J/\psi$ and $\psi(2S)$ charmonia.\\

\begin{figure}
    \includegraphics[width=.45\textwidth]{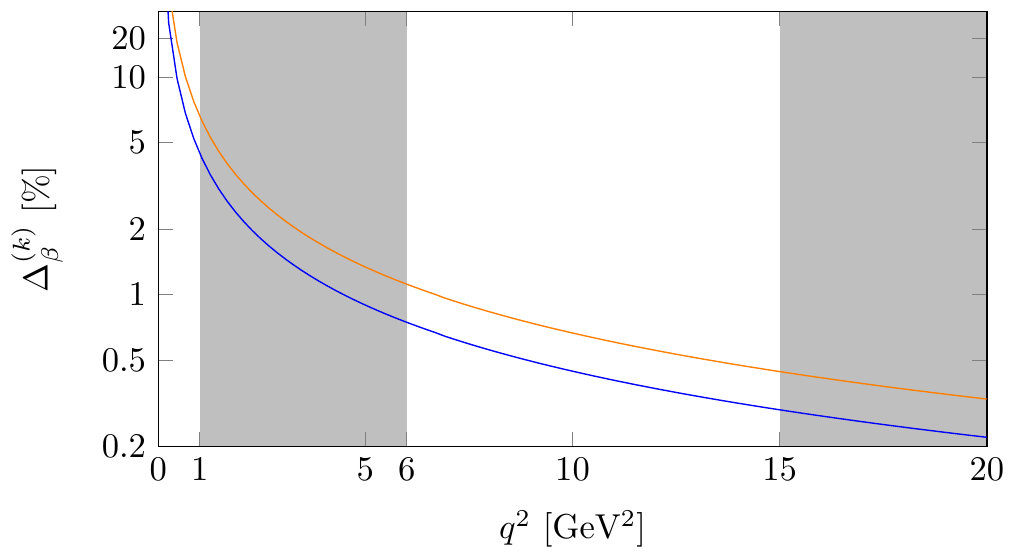}
    \caption{The $q^2$ dependence of the suppression terms $\Delta_\beta^{(k)}$
    for $k=2$ (lower blue curve) and $k=3$ (upper orange curve), respectively,
    on a log scale.  The windows in which either QCD-improved factorization or
    the operator product expansion at low recoil can be applied are indicated
    by the shaded area on the left and right, respectively.
    }
\end{figure}

For $q^2 \geq 1\,\GeV^2$, the suppressed terms in \refeq{results-Di} scale with
$\Delta_\beta^{(3)} < 6.6\%$ and $\Delta_\beta^{(2)} < 4.5\%$, respectively.
For the low recoil region, these terms further shrink down to $< 0.5\%$ and $<
0.3\%$, respectively.  Therefore, from a theoretical point of view, the low
recoil region would be ideal for our proposed analysis.  However, at LHCb the
experimental analysis of the $e^+e^-$ final state is more challenging for large
$q^2$.\\

While our approach is -- in principal -- also applicable to the angular
observables $S_i$ with $i=7,8,9$, we remind that any observation of
non-vanishing values for these observables already constitutes a sign of NP.\\

\begin{figure}
    \includegraphics[width=.5\textwidth]{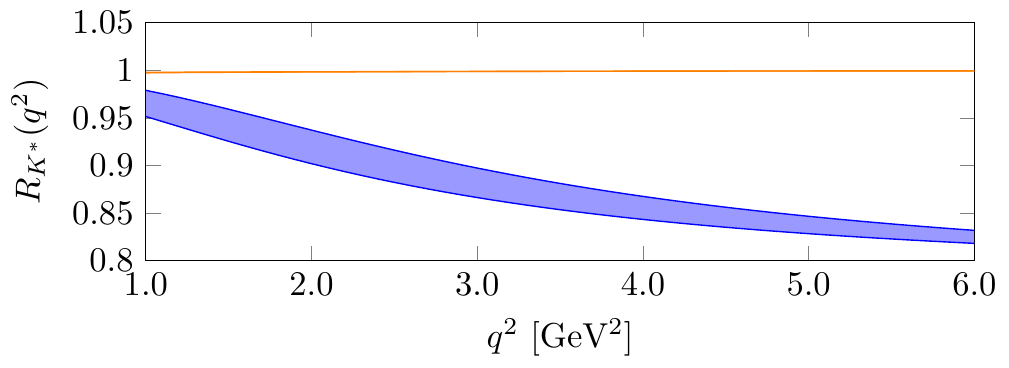}
    \caption{The $q^2$ behaviour of $R_{K^*}$ for our benchmark point
        $\wilson[$(e)$]{9} = \wilson[SM]{9} = \wilson[$(\mu)$]{9} + 1$
        (blue shaded area) and the SM (orange line). The illustrated uncertainty
        is due to our incomplete knowledge of the CKM matrix elements, the form factors,
        and the charm-induced long-distance contributions.
        The SM uncertainty too small to be shown.
    }
\end{figure}
Note that the proposed observables are not independent of $R_{K^*}$, the ratio of the decay rates into
$\mu$ versus $e$, since:
\begin{align}
    D_i(q^2) & = \left[\frac{\dd{\mathcal{B}^{(e)}}}{\dd q^2} + \frac{\dd{\mathcal{B}^{(e)}}}{\dd q^2}\right] \\
\nonumber    & \times \left[\omega^{(e)}(q^2) S^{(e)}_i(q^2) - \omega^{(\mu)}(q^2) S^{(\mu)}_i(q^2)\right]\,,
\end{align}
with $\mathcal{B}^{(\ell)}$ the branching ratio of $B\to K^*\ell^+\ell^-$, and weights
\begin{equation}
\label{eq:def-weights}
\begin{aligned}
    \omega^{(e)}(q^2)   & \equiv \frac{1}{1 + R_{K^*}(q^2)}\,, \\
    \omega^{(\mu)}(q^2) & \equiv \frac{R_{K^*}(q^2)}{1 + R_{K^*}(q^2)}\,. &
\end{aligned}
\end{equation}
Integration over $q^2$ from $a$ to $b$ then yields
\begin{multline}
    \int_{a}^{b} \dd q^2\,D_i(q^2) \equiv \langle D_i\rangle_{a,b}\\
    = \langle \mathcal{B}^{(e)}\rangle_{a,b}\,\langle S^{(e)}_i\rangle_{a,b} - \langle \mathcal{B}^{(\mu)}\rangle_{a,b}\,\langle S^{(\mu)}_i\rangle_{a,b}\,.
\end{multline}

We also wish to comment on opportunities for decays other than $B\to K^*(\to K\pi)\ell^+\ell^-$:
\begin{enumerate}
    \item The decays $B_s\to \phi(\to K^+K^-)\ell^+\ell^-$, with $\ell=e,\mu$
        can be described by the same angular PDF as $B\to K^*\ell^+\ell^-$
        decays. Thus a generalization of the $D_{i}$ observables to the
        $B_s$ decay is obvious. However, measurements of the
        theoretically most interesting observables $S_{5,6s}$ will require
        flavour tagging.
        Notice that the feasibility for this measurement in LHCb is both limited
        by the observed yield in Run-I and the low tagging power capability.
        Moreover, a flavour-tagged analysis at Belle II is very difficult, since
        the production of $B_s$ pairs at the $\Upsilon(5S)$ resonance does not
        occur through eigenstates of the charge-conjugation operators (unlike production
        of $B_d$ pairs at the $\Upsilon(4S)$).
    \item The observables $D_{i}$ can be generalized to the entire phase space
        of the $K\pi$ final state, i.e., to $K\pi$ masses outside the window
        that usually is associated with an on-shell $K^*(892)$. As for the
        $K^*(892)$, any significant deviation from zero, relative to the
        branching ratio, is a definite sign of LFU breaking, and thus a signal
        of NP. However, at the present time, the theoretical understanding of
        hadronic effects in $B\to K\pi\ell^+\ell^-$ is not well-enough
        developed for us to produce numerical estimates for small values of
        $q^2$.
    \item The decays $\Lambda_b\to \Lambda(\to N\pi)\ell^+\ell^-$ give rise to 10
        angular observables \cite{Boer:2014kda}. Amongst these observables,
        $K_{1c}$ and $K_{4s}$ permit a suppression of the charm-induced
        non-local matrix elements in the same fashion as shown in \refeq{results-Di}.
        However, at the present time, measurements of the muon final state are
        affected by large statistical uncertainties, and no measurements for
        the electron final state are available.
\end{enumerate}

\section{Numerical Results}
\label{sec:numerics}

In order to show that the observables $D_{4,5,6s}$ are indeed sensitive to LFU
breaking, we evaluate them at large hadronic recoil in one bin $1\,\GeV^2 \leq
q^2 \leq 6\,\GeV^2$, which we denoted as $\langle \cdot \rangle_{1,6}$. Our
numerical calculations are carried out using the EOS software \cite{EOS-Web},
which has been modified for this purpose \cite{EOS}. The evaluation of $B \to
K^*\ell^+\ell^-$ observables in the large recoil region implements the results
of the framework of QCD-improved factorization
\cite{Beneke:2001at,Beneke:2004dp}.  The uncertainties on the $D_i$ arise
dominantly from uncertainties of the CKM Wolfenstein parameters and our
incomplete knowledge of the $B \to K^*$ form factors.  The numerical input
values, their sources and their prior PDFs are listed in \reftab{inputs}. In
the SM, \ie, for $\wilson[$(e)$]{9} = \wilson[$(\mu)$]{9} = \wilson[SM]{9}$, we
obtain
\begin{equation}
\label{eq:results-Di-SM}
\begin{aligned}
    \langle R^\text{SM}_{K^*}\rangle_{1,6} & = 0.997 ^{+0.0005}_{-0.0004}\,,\\
    \langle D^\text{SM}_{4}  \rangle_{1,6} & = (+2.9^{+1.1}_{-1.7})\cdot 10^{-10}  &   &(^{+39}_{-60}\%)\,,\\
    \langle D^\text{SM}_{5}  \rangle_{1,6} & = (-1.1^{+2.0}_{-2.0})\cdot 10^{-10}  &   &(\pm 176\%)\,,\\
    \langle D^\text{SM}_{6s} \rangle_{1,6} & = (+4.4^{+1.7}_{-1.5})\cdot 10^{-10}  &   &(^{+40}_{-33}\%)\,.
\end{aligned}
\end{equation}
The large relative uncertainties in the SM are to be expected, since for
lepton-universal models the short-distance contributions on the right-hand side
of \refeq{results-Di} are small compared to the correction that involve the
hadronic contributions $h_{9,\lambda}$.

\begin{figure}
    \includegraphics[width=.49\textwidth]{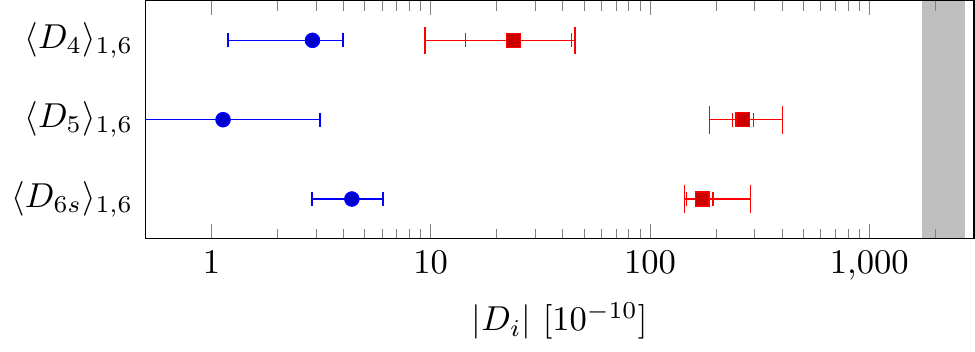}
    \caption{Comparison of the $q^2$-integrated observables $D_{i}$ in the SM
    (blue circles) with their values in the benchmark point (red squares) for
    the bin $1\,\GeV^2 \leq q^2 \leq 6\,\GeV^2$.  The smaller (red) error marks
    correspond to the pure form factor uncertainties, while the larger (red)
    error marks correspond to the uncertainties as obtained for the data driven
    scenario; see the text for more information. The experimental measurement
    of branching ratio $\mathcal{B}(B\to K^*\mu^+\mu^-)$, including the
    experimental uncertainty, is shown for comparison as the grey band.
    }
    \label{fig:comparison}
\end{figure}

However, in the case of LFU breaking, a sizeable leading short-distance term can
reduce the relative size of he non-local hadronic uncertainties. For comparison,
we define a benchmark point $\wilson[$(e)$]{9} = \wilson[SM]{9} =
\wilson[$(\mu)$]{9} + 1$.  This point is favoured by several global analyses of
data on $b\to s\ell^+\ell^-$ processes; see e.g.
\cite{Beaujean:2013soa,Altmannshofer:2014rta,Descotes-Genon:2015uva,Hurth:2016fbr}.
For our benchmark point (BMP) we obtain
\begin{equation}
\label{eq:results-Di-NP}
\begin{aligned}
    \langle R^\text{BMP}_{K^*}\rangle_{1,6} & = 0.86^{+0.02}_{-0.01}\,,\\
    \langle D^\text{BMP}_{4}  \rangle_{1,6} & = (+2.4^{+2.00}_{-0.95})\cdot 10^{-9} &    &(^{+84}_{-40}\%)\,,\\
    \langle D^\text{BMP}_{5}  \rangle_{1,6} & = (-2.7^{+0.31}_{-0.28})\cdot 10^{-8} &    &(^{+12}_{-11}\%)\,,\\
    \langle D^\text{BMP}_{6s} \rangle_{1,6} & = (-1.7^{+0.21}_{-0.26})\cdot 10^{-8} &    &(^{+12}_{-15}\%)\,,
\end{aligned}
\end{equation}
where the uncertainties of $D_{5,6s}$ are now dominated by the parametric CKM and form factor
uncertainties, while $D_{4}$ still shows large charm-induced uncertainties.
A comparison between \refeq{results-Di-SM} and \refeq{results-Di-NP} clearly
shows that the observables $D_{4,5,6s}$ are very sensitive to LFU-breaking NP
effects, with relative enhancements (for the central values only) of
\begin{equation}
\label{eq:enhancements-Di-BMP}
\begin{aligned}
    \frac{\langle D^\text{BMP}_{4} \rangle_{1,6}}{\langle D^\text{SM}_{4} \rangle_{1,6}} & \simeq 8\,,\\
    \frac{\langle D^\text{BMP}_{5} \rangle_{1,6}}{\langle D^\text{SM}_{5} \rangle_{1,6}} & \simeq 200\,,\\
    \frac{\langle D^\text{BMP}_{6s}\rangle_{1,6}}{\langle D^\text{SM}_{6s}\rangle_{1,6}} & \simeq 40\,.\\
\end{aligned}
\end{equation}
At the same time, it shows that the relative uncertainty is
reduced slightly for $D_4$, and strongly for $D_{5,6s}$.  This decrease in
(relative) uncertainty emerges, since the impact of the non-local hadronic
matrix elements is reduced compared to the now leading contributions from form
factors and $\wilson[$(e)$]{9} - \wilson[$(\mu)$]{9}$.\\

\begin{table}
    \renewcommand{\arraystretch}{1.2}
    \begin{tabular}{cccc}
        \hline
        Parameter                          & prior               & unit     & source\\
        \hline
        \multicolumn{4}{c}{CKM Wolfenstein parameters}\\
        \hline
        $\lambda$                          & $0.2253 \pm 0.0006$ & ---      & \cite{Bona:2006ah}\\
        $A$                                & $0.806 \pm 0.020$   & ---      & \cite{Bona:2006ah}\\
        $\bar{\rho}$                       & $0.132 \pm 0.049$   & ---      & \cite{Bona:2006ah}\\
        $\bar{\eta}$                       & $0.369 \pm 0.050$   & ---      & \cite{Bona:2006ah}\\
        \hline
        \multicolumn{4}{c}{Quark masses}\\
        \hline
        $\overline{m}_c(\overline{m}_c)$   & $1.275 \pm 0.025$   & \GeV     & \cite{Agashe:2014kda}\\
        $\overline{m}_b(\overline{m}_b)$   & $4.18 \pm 0.03$     & \GeV     & \cite{Agashe:2014kda}\\
        \hline
        \multicolumn{4}{c}{$B\to K^*$ power correction parameters}\\
        \hline
        $r_{0,\perp,\para}$                & $1.00 \pm 0.45$     & ---      & this work\\
        \hline
    \end{tabular}
    \renewcommand{\arraystretch}{1.0}
    \caption{%
        \label{tab:inputs}
        Numerical inputs for the calculations of the electron and muons
        components of the observables $D_i$. The power corrections $r_\chi$,
        $\chi=0,\perp,\para$ are scaling factors to the dipole form factors
        $\cal{T}$ as introduced in \cite{Beneke:2004dp}.  The prior
        distributions for all listed parameters are Gaussian, and the given
        intervals correspond to their central $68\%$ probability intervals.  We
        do not list the $B\to K^*$ form factor parameters here, which are taken
        from a simultaneous fit to Light-Cone Sum Rule and lattice QCD results
        \cite{Straub:2015ica}, including their correlation matrix. For the
        data-driven scenario, we also use $B\to K$ form factor parameters as
        obtained in \cite{Khodjamirian:2010vf}.
    }
\end{table}
We wish to further illustrate the usefulness of the newly-proposed observables
by studying a data-driven scenario (DDS). For this, we carry out a Bayesian fit involving a
free-floating $1.5 \leq \wilson[$(\mu)$]{9} \leq 5.5$, while we fix all other
Wilson coefficients to their SM values. The likelihood is comprised from the
LHCb measurement of $R_K$ \cite{Aaij:2014ora}, a recent preliminary result for $R_K$ by the
BaBar collaboration \cite{BaBarPreliminary}, as well as the LHCb results for $P'_5$ \cite{Aaij:2015oid},
an angular observables in the decay $B\to K^*\mu^+\mu^-$ that exhibits reduced
sensitivity to hadronic form factors.
We then proceed to produce posterior-predictive distributions for $R_{K^*}$ and
$D_{4,5,6s}$, which can be summarized as
\begin{equation}
\begin{aligned}
    \langle R^\text{DDS}_{K^*} \rangle_{1,6} & = 0.85 \pm 0.04\,,\\
    \langle D^\text{DDS}_{4}   \rangle_{1,6} & = ( 2.4^{+2.1}_{-1.5}) \cdot 10^{-9}\,,\\
    \langle D^\text{DDS}_{5}   \rangle_{1,6} & = (-3.1^{+0.9}_{-1.2}) \cdot 10^{-8}\,,\\
    \langle D^\text{DDS}_{6s}  \rangle_{1,6} & = (-2.1^{+0.8}_{-0.7}) \cdot 10^{-8}\,,
\end{aligned}
\end{equation}
which corresponds to qualitatively the same type of enhancements as in
\refeq{enhancements-Di-BMP}.  A comparison of all our numerical results for the
$D_{4,5,6s}$ is depicted in \reffig{comparison}.

\section{Experimental Feasibility}
\label{sec:exp}

\begin{figure}[t]
    \centering
    \includegraphics[width=.45\textwidth]{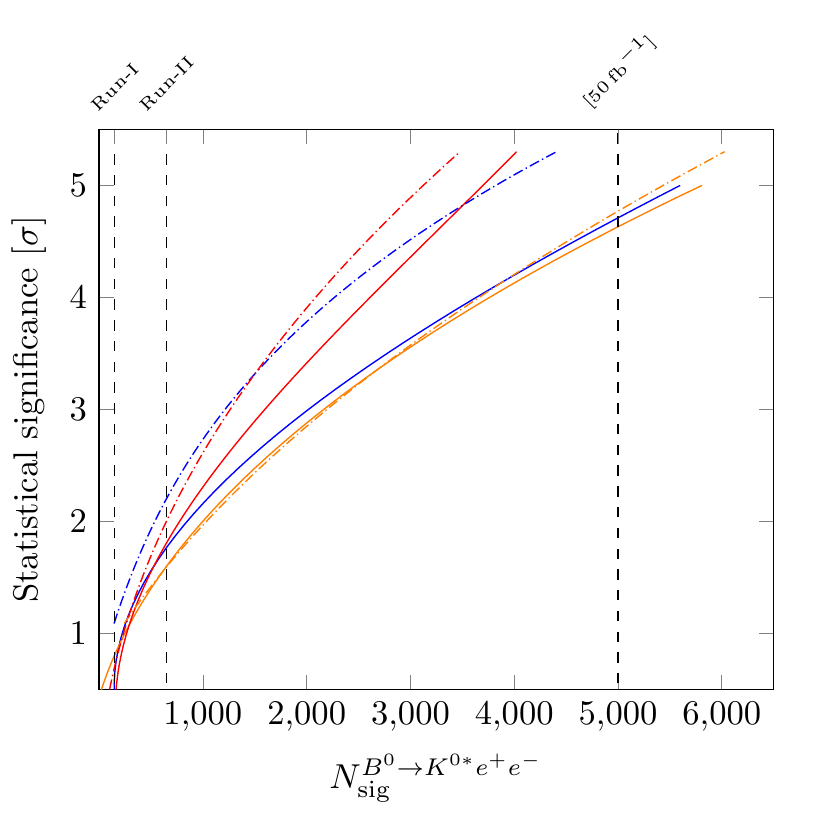}
    \caption{Projected statistical-only significance as a function of the
    extrapolated yields in LHCb for the (blue) $\langle D_{5} \rangle_{1,6}$
    and (orange) $\langle D_{6s}  \rangle_{1,6}$ observables, obtained from the
    (solid line) method-of-moments and (dash-dotted line) likelihood fit.  The
    red lines indicate the combined significance of $\langle D_{5}
    \rangle_{1,6}$ and $\langle D_{6s} \rangle_{1,6}$.
    }
    \label{fig:projection-Di}
\end{figure}

\begin{figure}[t]
    \centering
    \includegraphics[width=.45\textwidth]{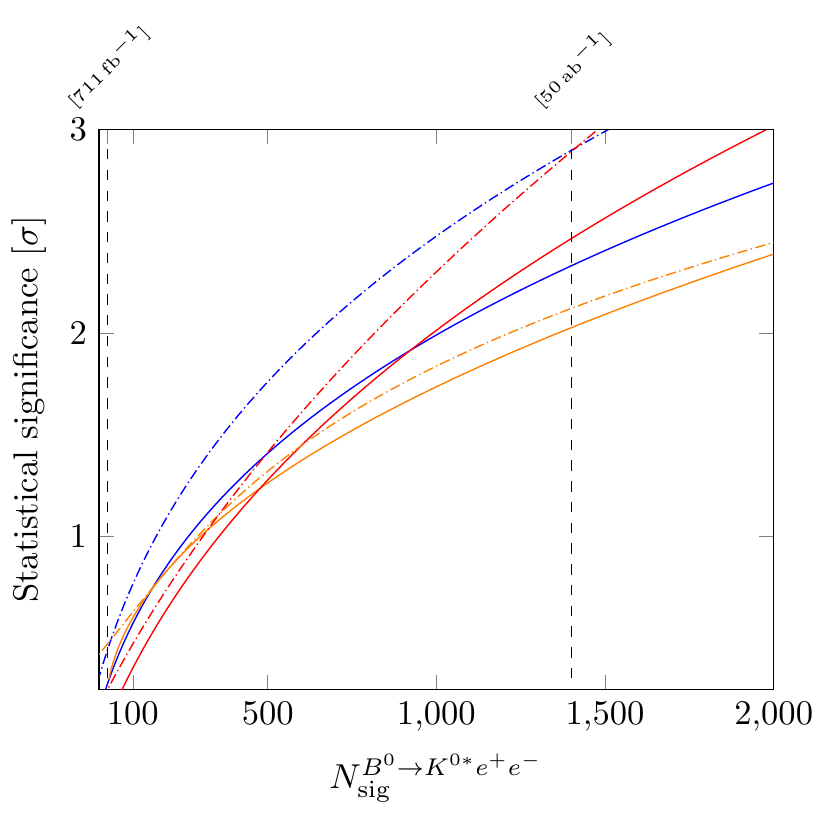}
    \caption{Projected statistical-only significance as a function of the
    extrapolated yields in Belle-II for the  (blue) $\langle D_{5}
    \rangle_{1,6}$ and (orange) $\langle D_{6s}  \rangle_{1,6}$ observables,
    obtained from the (solid line) method-of-moments and (dash-dotted line)
    likelihood fit.  The red lines indicate the combined significance of
    $\langle D_{5} \rangle_{1,6}$ and $\langle D_{6s} \rangle_{1,6}$.
    }
    \label{fig:projection-Di-Belle}
\end{figure}

A series of signal-only ensembles of pseudoexperiments is generated to
investigate the minimum amount of data required to claim an observation of NP
in these observables.  The simulation is performed without considering any
experimental effects, {\it{i.e.}}, background contributions, acceptance,
resolution or bin migration.  Similarly to the numerical calculations, the toy
model is implemented using the EOS framework independently for each final state
flavour at $1\,\GeV^2 \leq q^2 \leq 6\,\GeV^2$.  Pseudoexperiments are
generated with sample sizes corresponding roughly to the yields for the current
and forthcoming data taking periods available at LHCb and Belle
II.~\footnote{%
    Notice that effects of potential improvements ({\it{e.g.}} in the electron
    detection efficiency or reconstruction of $K^{*0}$ through $K_S \pi^0$),
    and of analysing additional signal channels ({\it{e.g.}} $B^+\to K^{*+}\ell^+\ell^-$)
    are not investigated here.
} These yields are extrapolated
(rounded to the nearest 50/500) from the values reported
in~\cite{Aaij:2014ora,Aaij:2015oid} and~\cite{Abdesselam:2016llu} by scaling
the luminosities and the $b\bar{b}$ cross section $\sigma_{b\bar{b}}$.  For
LHCb, we scale $\sigma_{b\bar{b}} \propto \sqrt{s}$, while for Belle
$\sigma_{b\bar{b}} \propto s$, where $s$ denotes the designed centre-of-mass
energy of the $b$-quark pair.  In particular, the significance for the range of
$[3\,$-$\,50]\,\invfb$ and $[1\,$-$\,50]\,{\rm{ab}}^{-1}$ are examined for LHCb
and Belle II, respectively.  Note that the relative yields between electrons
and muons are fixed in the pseudoexperiments.  Ensembles with other sample
sizes are also generated to test the scaling of the uncertainties, though only
a representative subset of the results obtained are shown here.

A convenient strategy to determine the $S^{({\ell})}_i(q^2)$, $\ell=e,\mu$
observables is to utilise the principal angular
moments~\cite{Beaujean:2015xea}.  Although this approach provides an
approximately $15\%$ worse precision on the measurement compared to the
likelihood fit~\cite{Aaij:2015oid},  as a proof-of-principle for the novel
observables this is more robust (e.g. against mismodelling of the PDF) and
insensitive to the choice of the estimated signal yield.  Nevertheless, for
completeness the results from an unbinned maximum likelihood fit are also
reported.  Note that the observables in both approaches correspond to the
average $\langle S^{({\ell})}_{i} \rangle_{1,6}$, obtained by summing over each
toy candidate for a given experiment.

Since the signal yield projections for the decay $B\to K^{*0} e^+ e^-$ in both
experiments indicate limited datasets, the stability of the likelihood fit is
enforced by simplifying the differential decay rate.  This is achieved by
applying folding techniques to specific regions of the three-dimensional
angular space, as detailed in Ref.~\cite{Aaij:2013qta}.  Notice that the
angular analysis is performed separately for each lepton flavour.  It is worth
emphasizing that, despite potential benefits on the experimental side, to
examine both final states simultaneously the choice to share/constrain angular
observables across different flavours should be in general avoided -- unless
otherwise strongly motivated. For instance, assuming $\wilson[$(e)$]{9} =
\wilson[SM]{9} = \wilson[$(\mu)$]{9} + 1$, $F_L^{(\mu)}$ is reduced by $10\%$
compared to $F_L^{(e)}$.\\

Furthermore, it has been shown that an alternative approach to weighting each
event by the inverse of its efficiency  is given by the calculation of the
unfolding matrix using the method of moments~\cite{Beaujean:2015xea}. This is
of particular interest in a simultaneous determination of the expressions
$\omega^{(\ell)}(q^2)$ (see \refeq{def-weights}) and $S^{(\ell)}_i(q^2)$, in
which a shared unfolded parametrisation can be used.  Further advantages
regarding the impact of common systematic effects are experiment dependent, and
therefore, not discussed in this note.

In order to obtain the profile of the statistical-only significance for the
extrapolated signal yields, both SM and NP simulated ensembles are examined.
The resulting $\langle D_{i} \rangle_{1,6}$ observables are fitted, and the
distance between the SM prediction and the fit results is calculated
in units of the toy measurements' standard deviations.
Additional constraints on the likelihood fit were necessary in order to ensure
the stability of the $\langle D_{6s} \rangle_{1,6}$ determination.  In
particular, the values of the longitudinal polarisation of the $K^{*0}$ meson
and the transverse polarisation asymmetry are constrained within uncertainties to
the theory predictions.

Figures~\ref{fig:projection-Di} and~\ref{fig:projection-Di-Belle} summarise the
expected sensitivity to benchmark-like NP effects for the proposed observables $\langle D_{5}
\rangle_{1,6}$ and $\langle D_{6s} \rangle_{1,6}$.  Projections for $\langle
D_{4} \rangle_{1,6}$ are not shown here since these are of limited usefulness.
Further studies with realistic experimental effects are necessary
to determine the exact sensitivities achievable.  An extrapolation of the
precision estimated here suggests that such measurements appear to be feasible,
albeit that the confirmation of benchmark-like NP effects independently
in each observable
is only possible with the full capability of the
experiments.  Furthermore, the correlation between these observables can be
estimated and used in a combined significance.
Based on our extrapolations, a first evidence of NP in the LHCb measurement and
considering only this novel approach can be achieved with $1300$ $B\to K^* e^+ e^-$ signal events,
which can currently not be expected to be recorded before the end of LHCb Run-II.
Similar sensitivity in the Belle-II measurement can be achieved with
$1500$ signal events, which corresponds roughly to an integrated luminosity of $50\,$ab$^{-1}$.
Note that the possibility of the proposed approach to go beyond the usual
theory upper bound of $q^{2}<6.0\,$GeV raises interesting prospects for
Belle-II: first, to increase their sensitivity due to stronger suppression
of the charm-induced contributions; and second, to record more events than
currently considered.

\section{Conclusion}
\label{sec:conclusion}

Recent measurements of $b\to s \ell^+\ell^-$ transitions show an interesting
pattern of deviations with respect to SM predictions. In particular, the
anomalous LHCb and Belle measurements of the observable $P_5^{\prime}$, and the
LHCb measurement of the LFU-probing ratio $R_K$ can
be simultaneously explained with NP contributions to the Wilson coefficients
$\wilson[$(\mu)$]{9}$ and/or $\wilson[$(\mu)$]{10}$.
This generated large attention in the flavour physics community, in particular
concerning long-distance charm-induced effects, which might be able to explain
the deviation in $P_5^{\prime}$.

Here, we proposed a new set of observables $D_i$ ($i=4,5,6s$), sensitive to
LFU-breaking effects in the decays $B\to K^* \ell^+\ell^-$. These observables
are branching-ratio-weighted averages of differences (with respect to the
final-state lepton flavour) of the angular observables $S_{4,5,6s}$.  In the
presence of the LFU-breaking NP effects in $\wilson[$(\ell)$]{9}$, their theoretical
uncertainties are dominated by $B\to K^*$ form factor and CKM parameter
uncertainties, while non-local hadronic contributions are kinematically
suppressed. This allows predictions in the NP scenarios that can be
systematically improved as our knowledge of the $B\to K^*$ form factors and CKM
Wolfenstein parameters improves. As examples we discussed here one benchmark
point, as well as a data-driven scenario based on a fit of the observable $R_K$
and $P'_5$.  All these scenarios have peculiar patterns of deviations of the
observables $D_i$ with respect to SM predictions (with reduced theoretical
uncertainties). Therefore these new observables, in addition to providing
sensitivity to discover NP with LFU-breaking effects, are useful to disentangle
the different scenarios and are crucial to test the mutual consistency across
different measurements.  It is important to highlight  that these observables
are also independent of other LFU-breaking measurements, {\it{e.g.}} $R_{K}$ or
$R_\phi$.  Hence, these can be included in global fits, which improves the
potential sensitivity to LFU-breaking NP effects.

The $D_i$ observables can be measured at the LHCb (and its upgrade) or at the
Belle II experiments, either by performing likelihood fits of the angular
distribution of the decays $B\to K^*\ell^+\ell^-$ or by using the method of
moments. We found that in order to obtain $3\sigma$ evidence for NP in only
these observables and using the method of moments, roughly 1500 $B\to K^* e^+ e^-$ signal events
are necessary in either experiment.

Our approach can be generalized for other decays to $K\pi \ell^+\ell^-$ final
states.  Here as well, a significant deviations from zero of the $D_i$
observables, relative in size to the branching ratio, would be a clear sign of
NP. However the theoretical and experimental knowledge of the $K\pi$ invariant
mass region outside the $K^{*0}(892)$ is not yet sufficient to provide solid
numerical predictions.

\acknowledgments

This work is supported by the Swiss National Science Foundation under grant PP00P2-144674.
We thank Marcin Chrzaszcz, Akimasa Ishikawa, Patrick Owen, Vincenzo Vagnoni, and Simon Wehle for
careful reading of the manuscript and valuable comments.
D.v.D.\ is grateful to the Mainz Institute for Theoretical Physics (MITP) for its
hospitality and its partial support during the completion of this work.

\FloatBarrier

\appendix

\section{Additional Formulae}

The full expressions for the observables $D_{4}$ through $D_{6s}$ in the basis of SM-like
operators, assuming real valued Wilson coefficients and LFU-breaking
only in the coefficient $\wilson{9}$, read:
\begin{widetext}
\begin{equation}
\label{eq:results-D4-full}
\begin{aligned}
    \frac{-D_{4}(q^2)}{\sqrt{2}\,N^2\,\tau_B}
        & = \left[\Delta_{9^2}^{(3)}\right] F_{V,\para} F_{V,0}
          - \Re\left[\Delta_9^{(3)} (h_{9,\para} + h_{9,0})\right] F_{V,\para} F_{V,0}\\
        & - \frac{2 m_b}{M_B} \Re\left[\Delta_9^{(3)} (\wilson{7} + h_{7,0})^*\right] F_{T,0} F_{V,\para}
          - \frac{2 m_b M_B}{q^2} \Re\left[\Delta_9^{(3)} (\wilson{7} + h_{7,\para})^*\right] F_{T,\para} F_{V,0}\\
        & + \Delta_\beta^{(3)} \frac{2 m_b^2}{s} \Re\left[(\wilson{7} + h_{7,0}) (\wilson{7} + h_{7,\perp})^*\right] F_{T,0} F_{T,\para}\\
        & + \Delta_\beta^{(3)} \left\lbrace \left|\wilson{10}\right|^2 + \Re\left[h_{9,0} h^*_{9,\para}\right]\right\rbrace F_{V,0} F_{V,\para}\\
        & + \Delta_\beta^{(3)} \frac{2 m_b M_B}{q^2} \Re\left[(\wilson{7} + h_{7,\para}) h^*_{9,0}\right] F_{T,\para} F_{V,0}
          + 3 \Delta_\beta^{(3)} \frac{2 m_b}{M_B}     \Re\left[(\wilson{7} + h_{7,0}) h^*_{9,\para}\right] F_{T,0} F_{V,\para}\,,
\end{aligned}
\end{equation}
and
\begin{equation}
\label{eq:results-D5-full}
\begin{aligned}
    \frac{D_{5}(q^2)}{2\,\sqrt{2}\,N^2\,\tau_B}
        & = 2 \Re\left[ \wilson{10} (\Delta_9^{(2)})^* \right] F_{V,\perp} F_{V,0}
          - \Delta_\beta^{(2)} \Re\left[\wilson{10} (h_{9,0} + h_{9,\perp})^*\right] F_{V,\perp} F_{V,0}\\
        & - \Delta_\beta^{(2)} \frac{2m_b}{M_B} \Re\left[\wilson{10} (\wilson{7} + h_{7,0})^*\right] F_{T,0} F_{V,\perp}
          - \Delta_\beta^{(2)} \frac{2m_b M_B}{q^2} \Re\left[\wilson{10} (\wilson{7} + h_{7,\perp})^*\right] F_{T,\perp} F_{V,0}\,,
\end{aligned}
\end{equation}
and
\begin{equation}
\label{eq:results-D6s-full}
\begin{aligned}
    \frac{D_{6s}(q^2)}{4\,N^2\,\tau_B}
        & = 2 \Re\left[ \wilson{10} (\Delta_9^{(2)})^*\right] F_{V,\para} F_{V,\perp}
          - \Delta_\beta^{(2)} \Re\left[\wilson{10} (\Delta_{9,\para} + \Delta_{9,\perp})^*\right] F_{V,\para} F_{V,\perp}\\
        & - \Delta_\beta^{(2)} \frac{2m_b M_B}{q^2} \Re\left[\wilson{10} (\wilson{7} + \Delta_{7,\perp})^*\right] F_{T,\perp} F_{V,\para}
          - \Delta_\beta^{(2)} \frac{2m_b M_B}{q^2} \Re\left[\wilson{10} (\wilson{7} + \Delta_{7,\para})^*\right] F_{T,\para} F_{V,\perp}\,.
\end{aligned}
\end{equation}
\end{widetext}

The form factors $F_{V,\lambda}$ and $F_{T,\lambda}$ for polarizations
$\lambda=0,\perp,\para$ are introduced \emph{ad hoc} in
\refeq{def-transamps}. The form factors for the vector and axialvector
currents are expressed as
\begin{align}
    F_{V,\perp}  & = \sqrt{2} \frac{\sqrt{\lambda}}{M_B + M_{K^*}} V\,,\\
    F_{V,\para}  & = \sqrt{2} (M_B + M_{K^*}) A_1\,,\\
    F_{V,0}      & = \frac{(M_B^2 - M_{K^*}^2 - q^2) (M_B + M_{K^*}) A_1}{2 M_{K^*} \sqrt{q^2}}\\
    \nonumber    & \quad - \frac{\lambda A_2}{2 M_{K^*} (M_B + M_{K^*}) \sqrt{q^2}}\,.
\end{align}
The form factors for the tensor current are expressed as
\begin{align}
    F_{T,\perp}  & = \sqrt{2} \frac{\sqrt{\lambda}}{M_B} T_1\,,\\
    F_{T,\para}  & = \sqrt{2} \frac{M_B^2 - M_{K^*}^2}{M_B} A_1\,,\\
    F_{T,0}      & = \frac{M_B (M_B^2 + 3 M_{K^*}^2 - q^2) T_1}{2 M_{K^*} \sqrt{q^2}}\\
    \nonumber    & \quad - \frac{\lambda T_1}{2 M_{K^*} (M_B^2 - M_{K^*}^2) \sqrt{q^2}}\,.
\end{align}
Here $V$, $A_{1,2}$ and $T_{1,2,3}$ are the form factors in the common
parametrization (see e.g. \cite{Bobeth:2012vn} for their definitions).

\bibliography{references}

\end{document}